\documentclass[traditabstract,a4,letter]{aa}  

\usepackage{epsfig}
\usepackage{graphicx}
\usepackage{epstopdf}
\usepackage{color}  
\usepackage{array}   
\usepackage{units}  
\usepackage[breaklinks,colorlinks, citecolor=blue]{hyperref}
\usepackage{natbib}  
\usepackage{multirow}   
\usepackage{amsmath,amssymb}  
\usepackage[english]{babel}
\usepackage[latin1]{inputenc}
\usepackage[T1]{fontenc}
\usepackage{longtable,lscape}
\usepackage{footnote}

\usepackage{txfonts}
\usepackage[toc,page]{appendix} 

\def\be{\begin{equation}}
\def\ee{\end{equation}}
\def\ba#1\ea{\begin{align}#1\end{align}}

\def\mr{\mathrm}

\definecolor{Mypink}{rgb}{1.0, 0.0, 0.5}

\begin{document}

\title{First detection of a virial shock with SZ data: \\ implication for the mass accretion rate of Abell 2319}
\author{G.~Hurier\inst{\ref{CEFCA}}, R.~Adam\inst{\ref{CEFCA}}, \and U.~Keshet\inst{\ref{afil2}}}

\institute{
  Centro de Estudios de F\'isica del Cosmos de Arag\'on (CEFCA), Plaza de San Juan, 1, planta 2, E-44001, Teruel, Spain 
  \label{CEFCA}
  \and
  Physics department, Ben-Gurion University of the Negev, POB 653, Be'er Sheva 84105, Israel 
  \label{afil2}
  \\ \\
  \email{ghurier@cefca.es}
}

\abstract{Shocks produced by the accretion of infalling gas in the outskirt of galaxy clusters are expected in the hierarchical structure formation scenario, as found in cosmological hydrodynamical simulations. Here, we report the detection of a shock front at a large radius in the pressure profile of the galaxy cluster A2319 at a significance of $8.6\sigma$, using Planck thermal Sunyaev-Zel'dovich data. The shock is located at $(2.93 \pm 0.05) \times R_{500}$ and is not dominated by any preferential radial direction. Using a parametric model of the pressure profile, we derive a lower limit on the Mach number of the infalling gas, $\mathcal{M} > 3.25$ at 95\% confidence level. These results are consistent with expectations derived from hydrodynamical simulations. Finally, we use the shock location to constrain the accretion rate of A2319 to $\dot{M} \simeq (1.4 \pm 0.4) \times 10^{14}$ M$_\odot$ Gyr$^{-1}$, for a total mass, $M_{200} \simeq 10^{15}$ M$_\odot$.}
\keywords{Cosmology: cosmic background radiation; Galaxies: clusters: intracluster medium; Galaxies: clusters: individual: A2319}
\authorrunning{G. Hurier {\it et al.}}
\titlerunning{Detection of Abell 2319 virial shock with {\it Planck} tSZ data}
\maketitle

\section{Introduction}
The structural properties of the hot ionized gas in galaxy clusters reflect their formation history through the merging of smaller substructures, but also from the continuous accretion of surrounding material from which they grow. On the large scales, the conversion from the infalling gas kinetic energy to thermal energy is expected to produce accretion shocks, as expected from hydrodynamical simulations \citep[see e.g.][]{Molnar2009}. Such shocks are excepted to produce a significant discontinuity in the pressure distribution of galaxy clusters around the virial radius, and generate high energy photons and cosmic rays via charged particle acceleration \citep[e.g.][]{bel78,bel78b,sar99,loe00,tot00,min01,kes03}. The properties of accretion shocks provide precious information about galaxy clusters mass accretion rate \citep[see e.g.][]{lau15}, and in addition, they can be used to infer the gas equation of state when combined with the galaxy clusters mass profile \citep[e.g.][]{shi16}.

Preliminary detection of accretion shocks in the outskirts of galaxy clusters have been achieved recently in the $\gamma$-ray in the case of the Coma cluster \citep{kes17a,kes17b} and for a stacked sample of galaxy clusters \citep{rei17}. However, the direct detection of the pressure drop in the intra-cluster medium (ICM) around the virial radius is particularly challenging. Indeed, X-ray observations, which play a fundamental role in the measurement of the ICM properties, are usually limited to the internal part of galaxy clusters \citep[$\lesssim R_{500}$,][]{planckppp} as the X-ray surface brightness is proportional to the gas density squared.

The thermal Sunyaev-Zel'dovich \citep[tSZ,][]{sun72} effect manifests as a spectral distortion of the Cosmic Microwave Background (CMB) blackbody radiation through the inverse Compton scattering on hot ionized electrons (a few keV) in the ICM \citep[see e.g.,][for reviews]{bir99,car02}. The tSZ effect offers a direct and linear probe of the electron pressure, $P_{\rm e}$, allowing to explore the intra-cluster gas up to large radius, providing that sufficient resolution and sensitivity is available. The tSZ is expected to show a cutoff at the virial radius as shown by \citet{koc05}. Its intensity is quantified by the Compton parameter, $y$, related to the integrated electronic pressure, $P_{\rm e}$, along the line of sight $d\ell$, as
\begin{equation}
y = \frac{\sigma_{{\rm T}}}{m_{\rm{e}} c^{2}} \int P_{\rm e} d\ell.
\label{tszeq}
\end{equation}
When expressed in units of CMB temperature, neglecting relativistic corrections \citep[see][for a detailed analysis of relativistic corrections]{hur17}, the tSZ signal is given by $\Delta T_{\rm{CMB}}/T_{\rm{CMB}}= g(\nu) \ y$.
where the CMB temperature is $T_{\rm CMB}$~=~2.726~K \citep{fix09}, and $g(\nu)$ the spectral dependence of the tSZ effect at the frequency $\nu$, 
\begin{equation}
g(\nu) = \left[ x\coth \left(\frac{x}{2}\right) - 4 \right] \quad \mr{with} \quad x=\frac{h \nu}{k_{\rm{B}} \, T_{\rm{CMB}}}.
\label{szspec}
\end{equation}
The tSZ signal is negative (positive) below (above) 217~GHz, independently of the cluster redshift \citep{hur14}.

During the last few years, a large number of tSZ experiments, both ground-based \citep[e.g.][]{act,nika,per15,spt,kit16} and satellite missions \citep{wmap,planck2015}, have proved the tSZ effect to be an excellent observable to trace shocks inside galaxy clusters \citep[see e.g.][]{planckcoma,nikafilter}. However, ground-based observations are limited in frequency coverage by the available atmospheric windows, such that the separation of the tSZ signal from other sky emissions (i.e. the CMB) is not always possible, which significantly increases the noise level on large scales. In addition, the recovered astrophysical signal generally suffers from the large scales filtering due to the observing strategy and the sky noise subtraction in the case of single dish, or the limited coverage of the uv plane for interferometers. Satellite-based mission are affected by their relatively low angular resolution \citep[$\sim$7' FWHM for {\it Planck},][]{planckbeam}, limiting the amount of resolved galaxy clusters in the sky. Due to these observational constraints, the detailed study of the pressure in clusters on very large scales remains challenging.

In the present paper we focus on the tSZ analysis of the outskirts of A2319, at redshift $z=0.056$ with a mass $M_{200}~=~(9.76 \pm 0.16)\times 10^{14}$ M$_\odot$ and a typical radius $R_{500}~=~1323 \pm 7$ Mpc \citep{ghi17}. This object is one of the brightest galaxy cluster observed by the {\it Planck} satellite mission via its tSZ signal and is well resolved \citep[$R_{500} \simeq 18'$,][]{mcxc}, allowing to search for pressure discontinuities at the cluster periphery. This cluster is also tought to be a major merger on smaller scales \citep{oha04}. In Sect.~\ref{sectmod}, we present the modeling of the pressure distribution of the cluster, including a shock component that is motivated by the discontinuity seen in the tSZ profile. The tSZ data processing and the fitting procedure are presented in Sect.~\ref{sectmeth}. In Sect.~\ref{sectconcl}, we derive constraints on the infalling gas Mach number and the mass accretion rate of A2319. Finally, we combine these results with the galaxy density profile of A2319 to derive constraints on the gas equation of state.

\section{Modeling of the pressure profile}\label{sectmod}
\begin{figure}[!h]
\begin{center}
\includegraphics[width=0.9\linewidth]{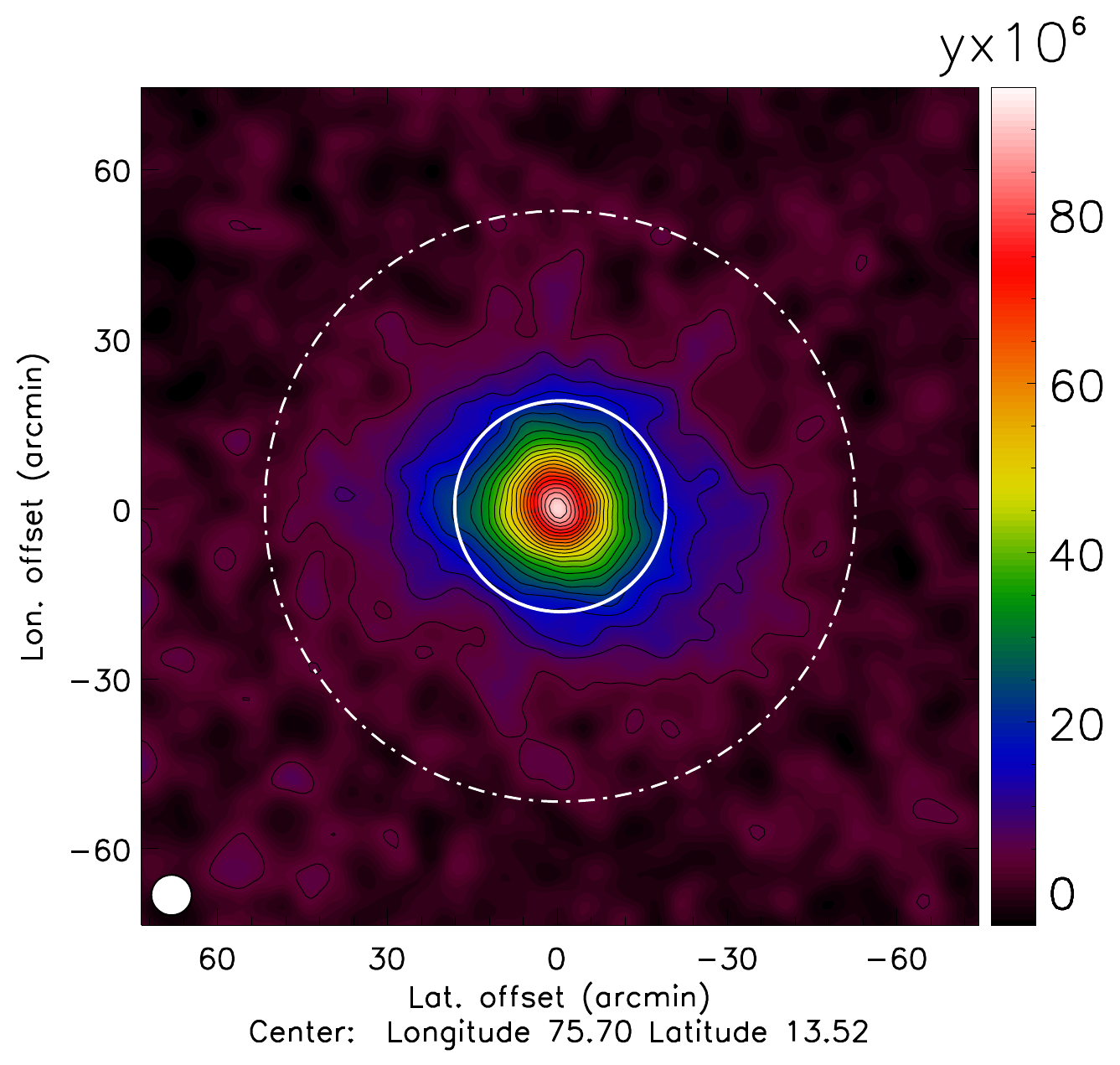}
\includegraphics[width=0.9\linewidth]{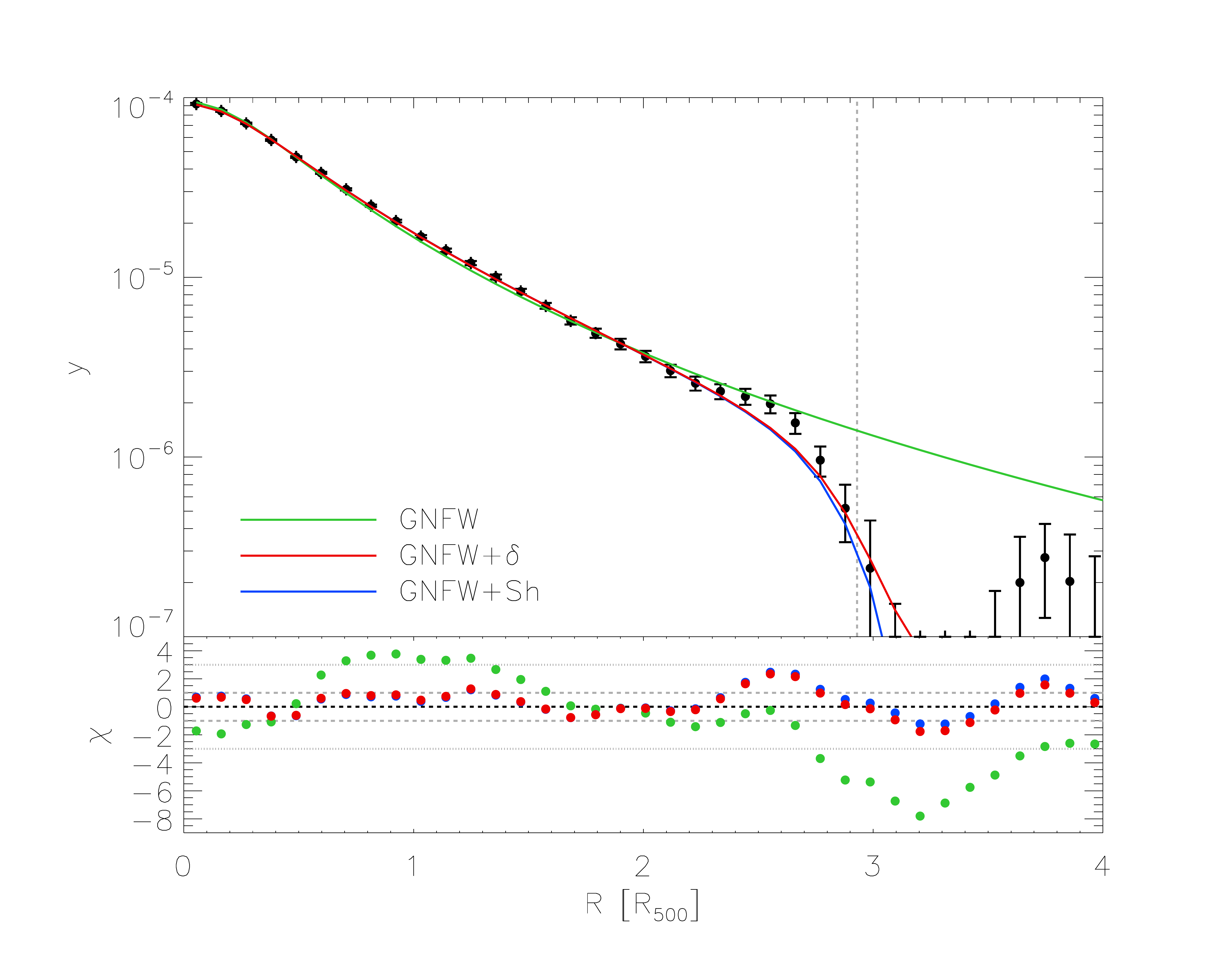}
\caption{\footnotesize {\bf Top:} MILCA tSZ map of A2319. The solid and dashed white circles show the typical radius of the cluster, $R_{500}$, and the shock radius, $R_{\rm Sh}$, detected in the azimuthal profile. The beam FWHM is given by the white circle on the bottom left and the black contours provide the signal to noise in units of $2\sigma$.
{\bf Bottom:} tSZ azimuthal profile of A2319. The black samples show the measured tSZ signal in the MILCA map, while the green, red and blue solid lines give the {\tt GNFW}, {\tt GNFW+$\delta$} (extra outer slope), and {\tt GNFW+Sh} (shock) best-fit models, respectively. The vertical dashed grey line corresponds to $R_{sh} = (2.93 \pm 0.05) \ R_{500}$. The residual, normalized by the error bars, $\chi$, is also shown for all the models.}
\label{a2319sz}
\end{center}
\end{figure}

In a hierarchical scenario of structure formation purely driven by gravitational collapse, the halo population is self-similar in scale and structure. The pressure profile of the galaxy clusters is well described by the parametric formulation given by \citet{gnfw}, referred to as the {\tt GNFW} model:
\begin{align}
\label{mod0}
P(x) = \frac{P_0 P_{500}}{(c_{500}x)^\gamma \left[ 1 + (c_{500}x)^\alpha \right]^{(\beta-\gamma)/\alpha}},
\end{align}
where $x = R/R_{500}$ is the scaled radius, and $P_{500}$ is the characteristic pressure. We refer to \citet{arn10} and \citet{planckppp} for previous fits to this parametric pressure profile description on a large sample of galaxy clusters using X-ray and tSZ data.

Motivated by the deviation observed from the {\tt GNFW} model for A2319 beyond $2.5\times R_{500}$ (see Sect.~\ref{sectmeth} and Fig.~\ref{a2319sz}), we also use two different models to account for a discontinuity in the pressure profile at large radii. First, we include a change in the outer slope (parameter $\delta$) of the pressure profile beyond $x_{\rm Sh} = R_{\rm Sh}/R_{500}$, and refer to this model as {\tt GNFW+$\delta$}:
\begin{align}
\label{mod1}
P(x) = \left\{
    \begin{array}{ll}
        \vspace{0.1cm}
        \frac{P_0 P_{500}}{(c_{500}x)^\gamma \left[ 1 + (c_{500}x)^\alpha \right]^{(\beta-\gamma)/\alpha}} & x < x_{\rm Sh} \\
	\frac{P_0 P_{500}}{(c_{500}x_{\rm Sh})^\gamma \left[ 1 + (c_{500}x_{\rm Sh})^\alpha \right]^{(\beta-\gamma)/\alpha}}\left(\frac{x}{x_{\rm Sh}}\right)^{-\delta} & x > x_{\rm Sh}
    \end{array}
\right.
\end{align}
Then, we use a {\tt GNFW} model, to which we add a jump in the pressure at radius $x_{\rm Sh}$. We parametrize the pressure drop with the parameter ${Q}_{\rm Sh}$. We refer to this model as {\tt GNFW+Sh}:
\begin{align}
\label{mod2}
P(x) = \left\{
    \begin{array}{ll}
        \vspace{0.1cm}
        \frac{P_0 P_{500}}{(c_{500}x)^\gamma \left[ 1 + (c_{500}x)^\alpha \right]^{(\beta-\gamma)/\alpha}} & x < x_{\rm Sh} \\
        \frac{P_0 P_{500}}{(c_{500}x_{\rm Sh})^\gamma \left[ 1 + (c_{500}x_{\rm Sh})^\alpha \right]^{(\beta-\gamma)/\alpha}}Q_{\rm Sh}\left(\frac{x}{x_{\rm Sh}}\right)^{-\delta} & x > x_{\rm Sh}
    \end{array}
\right.
\end{align}

While discrepancies between the {\tt GNFW} model and the data allow for the characterization of the significance of the pressure drop at large radii, a shock will be indicted by high values for $\delta$ in model {\tt GNFW+$\delta$}. Model {\tt GNFW+Sh} is physically motivated and allows us to describe the ICM properties, once a shock is clearly identified. From these pressure profile models, we compute the expected tSZ signal in {\it Planck} data by projecting the pressure profile on the line of sight according to Eq.~\ref{tszeq} and convolving the projected signal with a gaussian beam corresponding to the {\it Planck} tSZ angular resolution.


\begin{table*}
\center
\caption{\footnotesize Best fitting pressure profile parameters considering, {\it (i)} a GNFW profile, {\it (ii)} a GNFW profile + an outer slope, {\it (iii)} a GNFW profile + a shock. Only $\beta$ and $P_0$ are fitted for model {\tt GNFW}, and we fix $\alpha$, $\gamma$, and $c_{500}$ to the values derived in \cite{ghi17}, for the same cluster.}
\begin{tabular}{c|ccccc|ccc|cc}
Model names & $P_0$ & $c_{500}$ & $\alpha$ & $\beta$ & $\gamma$ & $R_{\rm Sh}$ [$R_{500}$] & ${Q}_{\rm Sh}$ & $\delta$ & $\chi^2_{\rm NDF}$ & NDF \\
\hline
{\tt GNFW} (Eq.~\ref{mod0}) & $9.68 \pm 0.17$ & $1.10$ & $1.05$ & $4.552 \pm 0.033$ & $0.23$ & / & / & / & 6.0 & 35 \\
{\tt GNFW+$\delta$} (Eq.~\ref{mod1}) & $8.83 \pm 0.16$ & $1.10$ & $1.05$ & $4.304 \pm 0.033$ & $0.23$ & $2.93 \pm 0.06$ & / & > 84 & 1.2 & 33 \\
{\tt GNFW+Sh} (Eq.~\ref{mod2}) & $8.82 \pm 0.20$ & $1.10$ & $1.05$ & $4.295 \pm 0.038$ & $0.23$ & $2.93 \pm 0.05$ & $ -0.012 \pm 0.045$ & $\beta$ & 1.1 & 33  \\
\end{tabular}
\label{tabgnfw}
\end{table*}

\section{Methodology}\label{sectmeth}
\begin{figure}[!h]
\begin{center}
\includegraphics[width=1.0\linewidth]{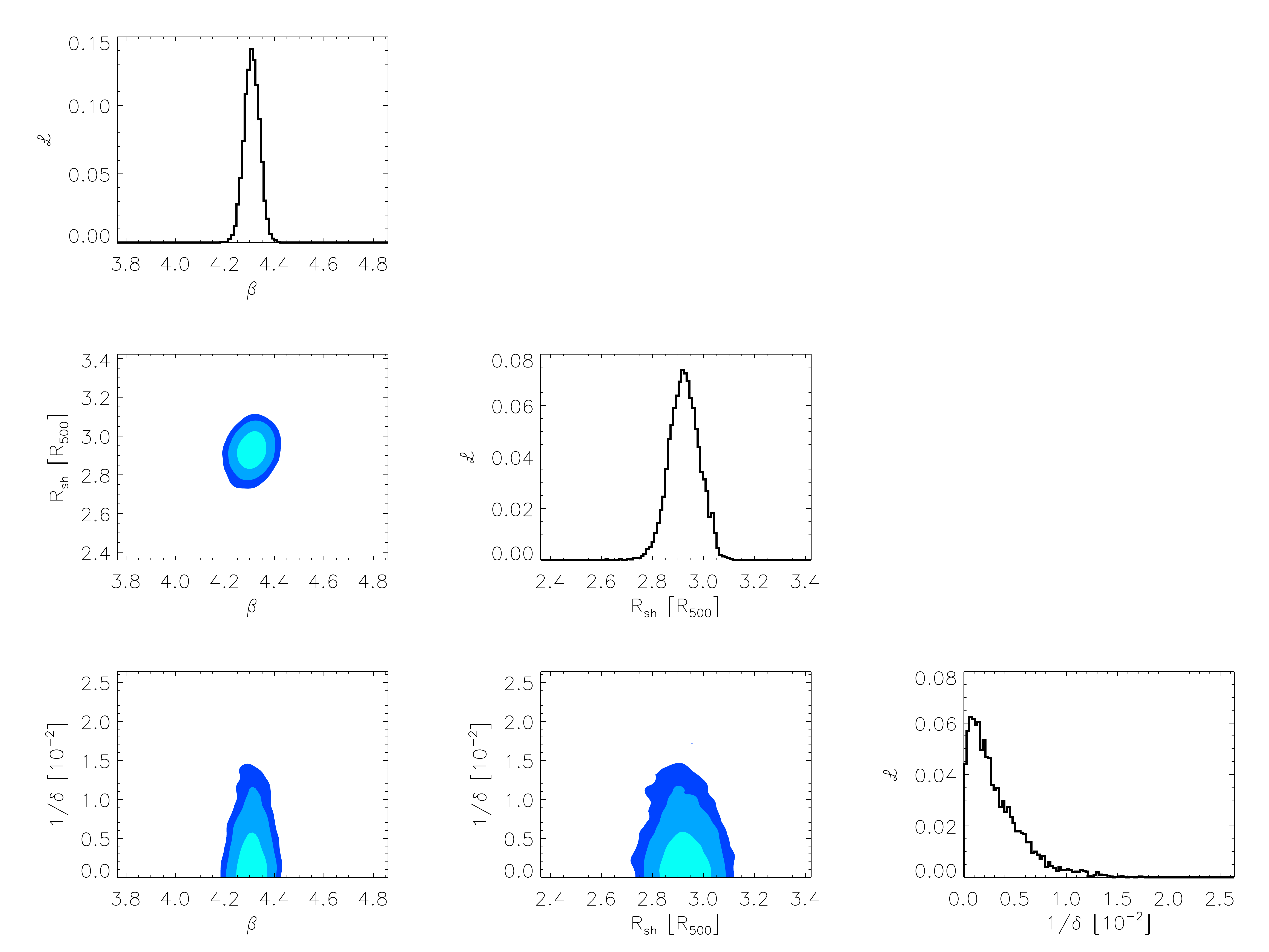}
\caption{\footnotesize Posterior likelihood functions for model {\tt GNFW+$\delta$}. Histograms on the diagonals present the marginalized likelihoods for parameters $\beta$, $R_{\rm Sh}$, and $1/\delta$. The others panels show the marginalized likelihood with the light blue, blue, and dark blue contours providing confidence levels at 68.3, 95.4, and 99.7\%, respectively.}
\label{a2319like}
\end{center}
\end{figure}

The core of our analysis consists of characterizing the drop of pressure, as seen through the tSZ effect, in the outskirt of A2319. First, we construct a tSZ map of A2319 at 7' FWHM angular resolution using MILCA \citep{hur13} and {\it Planck} data \citep{planck2015}. This map is derived from a pixel and scale dependent linear combination of {\it Planck} frequency channels from 100 to 857 GHz \citep[see also][for more details]{hur15}. We verify that using the {\it Planck} public tSZ maps at a resolution of 10' FWHM \citep[MILCA \& NILC,][]{plancksz} provides consistent results. In the top panel of Fig.~\ref{a2319sz}, we present the tSZ map of A2319. We observe a clear tSZ signal from A2319, with a central value $y_0 \simeq 10^{-4}$, reaching more than 45$\sigma$ per beam, at this resolution. We note that the tSZ map angular resolution (shown as a white circle in Fig.~\ref{a2319sz}) is small compared to the projected characteristic radius of A2319, $R_{500} = 18.4$' \citep{mcxc}.

Then, we compute an azimuthally averaged tSZ profile using concentric annuli with a 2' binning ($\simeq 9$ samples per $R_{500}$). By construction, the noise in the MILCA tSZ map at 7' FWHM has a typical correlation length of 5' FWHM. It is thus mandatory to account for the full covariance matrix of the tSZ profile. Due to it's scanning strategy the noise in {\it Planck} data is inhomogeneous \citep{planck2015}. Thus, we estimated the covariance matrix, ${\cal C}_{P}$, of the tSZ profile using 1000 simulations of inhomogeneous correlated gaussian noise \citep[see][for a description of the noise simulation procedure]{planckppp}. We present the extracted tSZ profile and the associated uncertainties, up to $4 \times R_{500}$, in the bottom panel of Fig.~\ref{a2319sz}. We observe a clear drop of the tSZ signal around $R \simeq 2.8 \times R_{500}$. We check that this drop is not dominated by any preferential direction and is consistent among several independent profiles extracted from different slices in the map.

We adjust the models presented in Sect.~\ref{sectmod} to the measured tSZ profile using a Markov Chain Monte Carlo approach and assuming that the uncertainties follow gaussian statistic, using
\begin{align}
\chi^2 = \left(\widehat{{\bf P}} - {\bf P} \right) {\cal C}^{-1}_{P} \left( \widehat{{\bf P}} - {\bf P} \right)^{\rm T},
\end{align}
where $\widehat{{\bf P}}$ is the measured tSZ profile and ${\bf P}$ the tSZ profile derived from the models. The best fitting parameters are listed in Table~\ref{tabgnfw} for the three pressure profile models that we consider: a GNFW pressure profile ({\tt GNFW}), a GNFW pressure profile with a change of slope at $R = R_{\rm Sh}$ ({\tt GNFW+$\delta$}), and a GNFW pressure profile with pressure drop at $R = R_{\rm Sh}$ ({\tt GNFW+Sh}). The present analysis focusses on the large scale behavior of the pressure profile of A2319, and we thus fix the parameters $\alpha$, $\gamma$, and $c_{500}$ to the values derived in \citet{ghi17} because they are related to the inner structure of the cluster. We fit for the parameters $\beta$ and $P_0$ in the case of model {\tt GNFW}, $\beta$, $P_0$, $R_{\rm Sh}$, and $1/\delta$ in the case of model {\tt GNFW+$\delta$}, and $\beta$, $P_0$, $R_{\rm Sh}$, and ${Q}_{\rm Sh}$ for model {\tt GNFW+Sh}, for which we fixed $\delta = \beta$.

Models {\tt GNFW+$\delta$} and {\tt GNFW+Sh} provide a good description of the observed profile at all scales with a reduced $\chi^2_{\rm NDF}$ of 1.2 and 1.1, respectively. Model {\tt GNFW} fails to describe the data with a $\chi^2_{\rm NDF} = 6.0$, and model {\tt GNFW+Sh} is favored by $8.5\, \sigma$ compared to model {\tt GNFW}. 
In Fig.~\ref{a2319like}, we present the posterior likelihood function on the fitted parameters for model {\tt GNFW+$\delta$}. 
The outer slope parameters is constrained to $\delta > 84$ at 99\% confidence level. This indicates that the tSZ drop located at $R_{\rm Sh}$ is due to a discontinuity in the pressure, corresponding to a shock in the ICM. In contrast, the slope of the pressure profile at radius $R \lesssim R_{\rm Sh}$ is constrained to $\beta = 4.30 \pm 0.03$. Model {\tt GNFW+Sh} shows no significant pressure detected beyond $R_{\rm Sh}$, with $Q_{\rm Sh} = -0.012 \pm 0.045$. We note that this results does not depends on the chosen value for $\delta$ in the {\tt GNFW+Sh} model.

\section{Results and conclusion}\label{sectconcl}
\begin{figure}[!h]
\begin{center}
\includegraphics[width=0.9\linewidth]{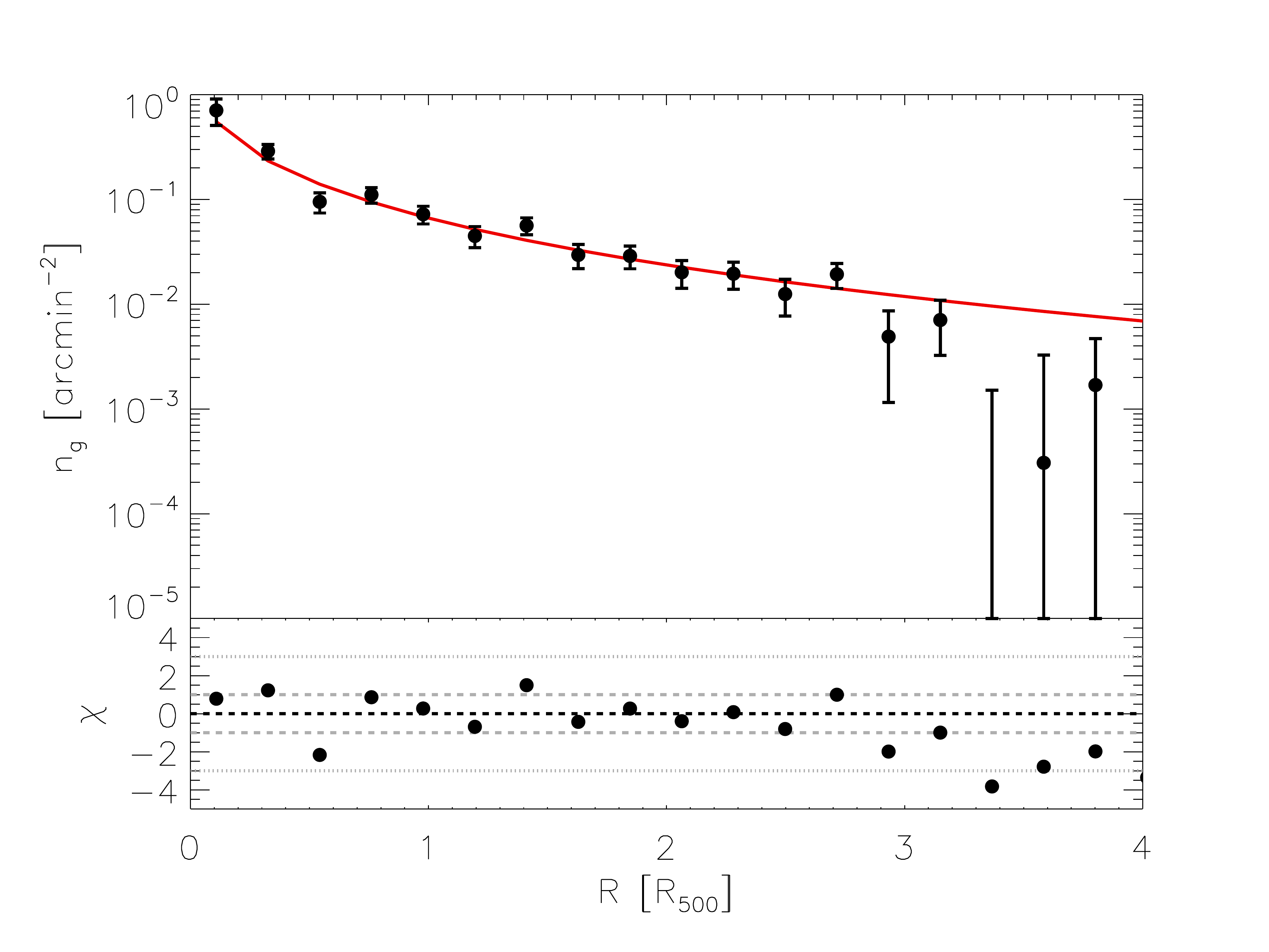}
\includegraphics[width=0.9\linewidth]{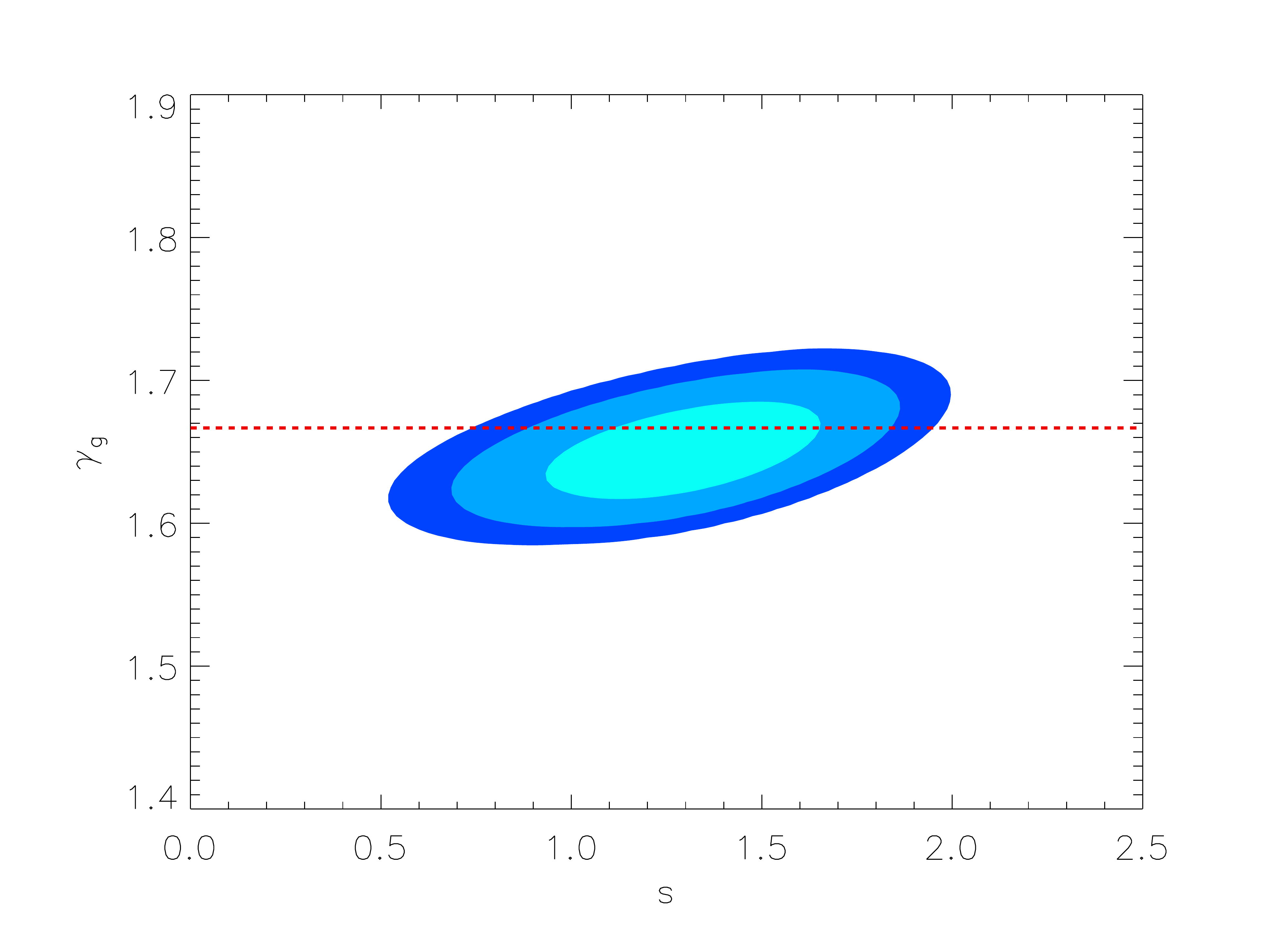}
\caption{\footnotesize {\bf Top:} Galaxy density profile of A2319 derived from 2MASS catalog. Back samples show the data and the solid red line shows the best fitting NFW profile for $R < 2.5 \times R_{500}$. 
{\bf Bottom:} Accretion rate and equation of state likelihood.}
\label{a2319sg}
\end{center}
\end{figure}

The amplitude of the pressure discountinuity, $Q_{\rm Sh}$, is related to the infalling gas Mach number \citep{lan59,sar02}, $\mathcal{M}$, as
\begin{align}
Q_{\rm Sh} = 2 \frac{\gamma_g}{\gamma_g + 1}{\mathcal{M}}^2 - \frac{\gamma_g-1}{\gamma_g+1},
\end{align}
where $\gamma_g$ is the gas adiabatic index (we expect $\gamma_g = 5/3$ for a fully ionized plasma). In the case of A2319, no significant tSZ signal is observed at $R > R_{\rm Sh}$ and thus, only a lower limit can be set on the Mach number, constrained to $\mathcal{M} > 3.25$ at 95\% confidence level. Following the results by \citet{shi16}, we can use the location of the virial shock to constrain the accretion rate of the galaxy cluster. Assuming $\gamma_g~=~5/3$ and self similarity, we obtain, $s=~1.36~\pm~0.20$, corresponding to the cluster mass evolving as $M(R) \propto t^{3s/2}$.

Finally, we compare the virial shock radius to the galaxy number density profile of A2319, which we extract using the 2MASS \citep{mass} galaxy survey (see top panel of Fig.~\ref{a2319sg}). The galaxy number density profile is well described by a NFW profile \citep{nfw} below $2.8 \times R_{500}$, with a concentration $c_g = 1.00 \pm 0.37$. At radii $R > 3 \times R_{500}$ we observe a $>4 \, \sigma$ deficit in the galaxy number density profile compared to a NFW profile extrapolation. Assuming that this drop corresponds to the splashback radius, $R_{\rm sp} = \left(3.2 \pm 0.1\right) R_{500}$, we use the results from \citet{shi16} to constrain the gas adiabatic index, in combination with the location of the virial shock. We present the corresponding posterior likelihood in the plane $s$ -- $\gamma_g$ in the bottom panel of Fig.~\ref{a2319sg} and constrain the parameters to $s = 1.29 \pm 0.24$ and $\gamma_g = 1.65 \pm 0.02$. This result is consistent with the expectation of $\gamma_g = 5/3$, and allows us to measure the accretion rate of A2319, $\dot{M} \simeq (1.4 \pm 0.4) \times 10^{14}$~M$_\odot$Gyr$^{-1}$, in agreement with expectation from numerical simulation considering A2319 mass \citep[see e.g.,][]{deb16}.
This results illustrates the strength of high signal-to-noise large scale tSZ observations to constraints galaxy cluster formation by directly probing the outskirts of galaxy clusters where the baryons are heated.

\section*{Acknowledgment}
\thanks{\footnotesize
We thanks R. Angulo, C. Hern\'andez-Monteagudo, and O. Hahn for useful discussions. 
G.H. and R.A, acknowledge support from Spanish Ministerio de Econom\'ia and Competitividad (MINECO) through grant number AYA2015-66211-C2-2. U.K. acknowledges support by the Israel Science Foundation (ISF grant No. 1769/15).
We acknowledge the use of HEALPix \citep{gor05}.
}

\bibliographystyle{aa}
\bibliography{virial_shock}

\end{document}